\documentclass[aps, prb, twocolumn,10pt,longbibliography, superscriptaddress]{revtex4-1}

\usepackage{amsmath, amssymb, amsfonts, amsthm, bm, mathrsfs, bbm}
\usepackage{graphicx}
\usepackage{epstopdf}
\usepackage{color}
\usepackage[usenames,dvipsnames]{xcolor}
\usepackage[citecolor=blue, colorlinks=true, urlcolor=blue, linkcolor=blue]{hyperref}

\newcommand{\eq}[1]{Eq.~(\ref{#1})}

\newcommand{\fig}[1]{Fig.\thinspace{}\ref{#1}}

\newcommand{\fc}[1]{({#1})}

\usepackage{times}

\begin{document}

\title{Dynamical Quantum Phase Transitions in Systems with Continuous Symmetry Breaking}

\author{Simon A. Weidinger}
\email[]{simon.weidinger@tum.de}
\affiliation{Department of Physics and Institute for Advanced Study, Technical University of Munich, 85748 Garching, Germany}

\author{Markus Heyl}
\affiliation{Max-Planck-Institut f\"ur Physik komplexer Systeme, 01187 Dresden, Germany}
\affiliation{Department of Physics and Institute for Advanced Study, Technical University of Munich, 85748 Garching, Germany}

\author{Alessandro Silva}
\affiliation{SISSA - International School for Advanced Studies, via Bonomea 265, 34136 Trieste, Italy}

\author{Michael Knap}
\affiliation{Department of Physics and Institute for Advanced Study, Technical University of Munich, 85748 Garching, Germany}

\begin{abstract}

Interacting many-body systems that are driven far away from equilibrium can exhibit phase transitions between dynamically emerging quantum phases, which manifest as singularities in the Loschmidt echo. Whether and under which conditions such dynamical transitions occur in higher-dimensional systems with spontaneously broken continuous symmetries is largely elusive thus far. Here, we study the dynamics of the Loschmidt echo in the three dimensional O(N) model following a quantum quench from a symmetry breaking initial state. The O(N) model exhibits a dynamical transition in the asymptotic steady state, separating two phases with a finite and vanishing order parameter, that is associated with the broken symmetry. We analytically calculate the rate function of the Loschmidt echo and find that it exhibits periodic kink singularities when this dynamical steady-state transition is crossed. The singularities arise exactly at the zero-crossings of the oscillating order parameter. As a consequence, the appearance of the kink singularities in the transient dynamics is directly linked to a dynamical transition in the order parameter.  Furthermore, we argue, that our results for dynamical quantum phase transitions in the O(N) model are general and apply to generic systems with continuous symmetry breaking.

\end{abstract}

\date{\today}

\pacs{
}

\maketitle



\section{Introduction}
\label{sec:Intro}

In recent years, synthetic quantum matter such as ultra-cold atoms, polar molecules, and trapped ions have demonstrated their capabilities to experimentally study nonequilibrium quantum states far beyond the regime of linear response and thus far beyond a thermodynamic description. Due to the isolation from the environment and the high level of control, experiments with synthetic quantum matter have shown that inherently dynamical phenomena can be realized and probed, ranging from many-body localization,\cite{Schreiber15, Choi16, Smith16, Bordia16, Bordia17, Luschen16, Bordia17b}, prethermalization,\cite{Gring12, Langen15} discrete time crystals,\cite{Monroe17, ChoiS17} the particle-antiparticle production in the Schwinger model,\cite{Martinez16} to emergent Bloch oscillations.\cite{Meinert17}  In addition, not only the dynamical phases themselves have become accessible in experiments, but also the associated dynamical transitions between the phases.\cite{Jurcevic16, Flaschner2016, Zhang17}

Current experimental platforms for studying dynamics are often focusing on one- and two-dimensional systems. Yet, a future prospect concerns extensions toward the realization of non-equilibrium many-body states in three spatial dimensions, where new physical phenomena become accessible. This includes, for example, the possibility of spontaneously broken continuous symmetries at nonzero temperatures, which is excluded for lower dimensions due to the Mermin-Wagner theorem in systems with short range interactions.

\begin{figure}
	\includegraphics[width=.46\textwidth]{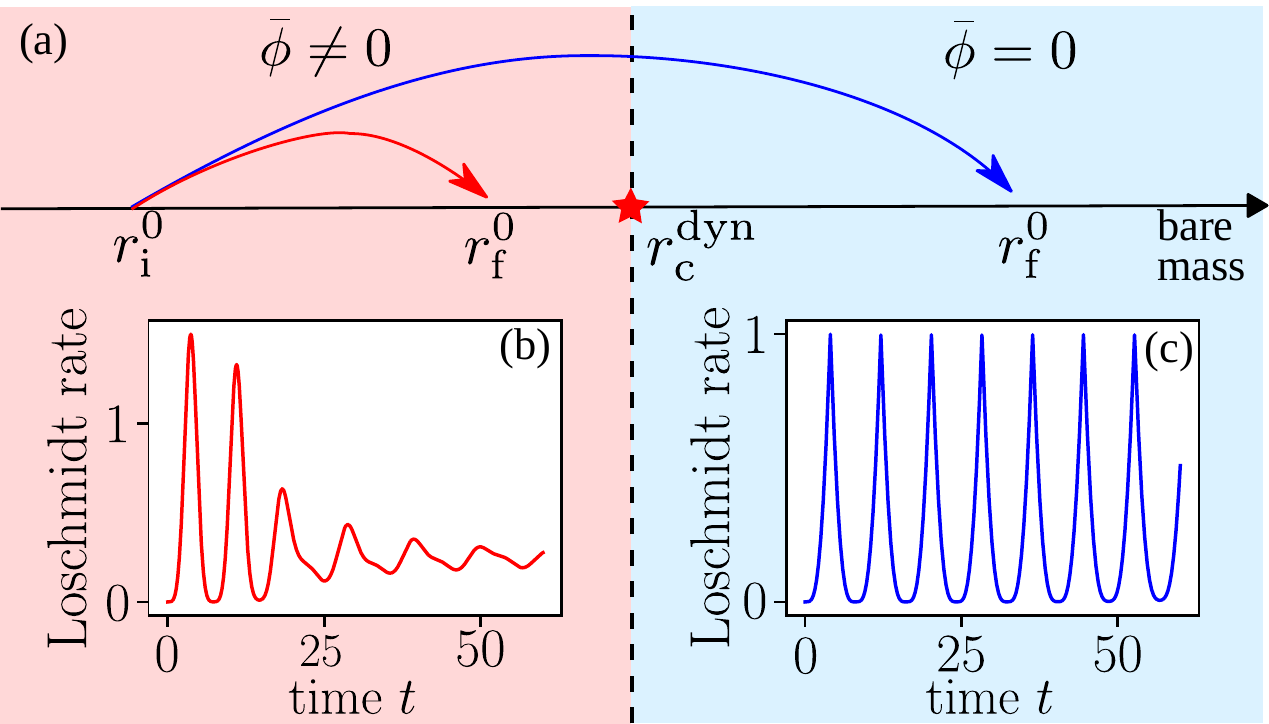}
	\caption{\textbf{Dynamical criticality in the Loschmidt echo for systems with spontaneous symmetry breaking}. \fc{a} We study dynamical quantum phase transitions of the O(N) model following quantum quenches from an initial bare mass $r_i^0$ to a final bare mass $r_f^0$. The initial state is chosen to break the continuous symmetry of the O(N) model and hence is described by a finite order parameter. Our system exhibits a steady-state dynamical phase transition at $r_c^\text{dyn}$, which separates the dynamically ordered phase in which the long-time average of the order parameter $\bar \phi$ remains finite, from the disordered phase in which $\bar \phi$ vanishes. We analytically calculate the Loschmidt echo and find that the associated rate function  remains smooth for  quenches within the dynamically ordered phase \fc{b} but  exhibits nonanalytic kink singularities when crossing the dynamical critical point $r_c^\text{dyn}$ \fc{c}. }
	\label{fig:schematic}
\end{figure}

In this work, we study the quantum dynamics of an interacting many-body system in three dimensions which exhibits such a spontaneously broken symmetry. Specifically, we calculate the unitary real-time evolution of the O(N) vector model following a quantum quench of the mass, with an initial state that breaks the continuous symmetry of our system, see \fig{fig:schematic}. We approach the problem fully analytically via the large-$N$ limit, where the dynamics can be solved exactly. 

The O(N) model exhibits a dynamical quantum phase transition in the asymptotic steady state, separating two dynamical phases with finite and vanishing order parameter, respectively.~\cite{Sciolla:2013aa} Here, we show that in addition to the dynamical steady-state transition of the order parameter, the O(N) model exhibits a critical dynamical phenomenon on transient time scales. In particular, non-analyticities appear in the Loschmidt echo periodically in time when the dynamical transition is crossed by the quantum quench (\fig{fig:schematic}). We show that in the O(N) model these singularities contribute only subextensively to the rate function associated with the Loschmidt echo. Making use of the analogy between the Loschmidt echo and the boundary partition function, this effect is reminiscent of surface phase transitions in equilibrium systems, which also contribute only subextensively to the free energy.\cite{Diehl97} Furthermore, we find that the dynamical critical point obtained from the order parameter coincides with the one obtained from the Loschmidt echo. These different concepts of dynamical criticality are further linked by the fact that the non-analyticities in the Loschmidt echo occur at  times when the order parameter crosses zero. A similar relation has been found in the long-range transverse-field Ising model.\cite{Zunkovic2016} We argue that our results are not specific to the O(N) model or the large-$N$ limit, and hence apply to generic systems with a spontaneously broken continuous symmetry.

This work is organized as follows. In Sec.~\ref{sec:DQPTs} we discuss two different notions of dynamical quantum phase transitions; one associated with the time evolution of the order parameter, and the other one exploiting the similarity between the Loschmidt echo and a classical partition function. In Sec.~\ref{sec:ONmodel} we review the far from equilibrium dynamics of the $O(N)$ model in the large $N$ approximation to leading order. The time evolved state of the $O(N)$ model is calculated in Sec.~\ref{sec:state}. In Sec.~\ref{sec:returnprob} we derive the return probability of the time evolved state to the ground state manifold and in Sec.~\ref{sec:ratefunction} we analyze the corresponding rate function and show that it exhibits non-analyticities, when the model is quenched across the dynamical critical point. Finally, in Sec.~\ref{sec:outlook} we conclude our findings and discuss potential extensions of our work.

\section{Dynamical Quantum Phase Transitions}
\label{sec:DQPTs}

We investigate two notions of dynamical quantum phase transitions.
The first one, is associated with the time evolution of the order parameter.\cite{Yuzbashyan2006, Eckstein2010, Gambassi2010, Schiro2010, Sciolla2011, Sciolla:2013aa, Hamerla2013, Smacchia:2015aa, Zunkovic2016} The dynamical quantum phase transition is then characterized by a critical point which separates regimes where the long-time average of the order parameter $\bar \phi$ is either finite or zero. Close to this dynamical critical point the long-time average $\bar \phi$ exhibits scaling relations with critical exponents.\cite{Sciolla:2013aa, Smacchia:2015aa} However, the location of the dynamical critical point can in general differ from the equilibrium one and might also depend on the initial state, due to a dynamical renormalization of parameters.\cite{Smacchia:2015aa, Sciolla:2013aa}

A second approach to study the nonequilibrium dynamical criticality is to exploit the formal similarity between the equilibrium partition function $Z = \mathrm{tr} [e^{-\beta \hat{H}}]$ and the Loschmidt amplitude $\langle\psi_0|e^{-i \hat{H}t}|\psi_0\rangle$. \cite{Heyl:2013aa, Heyl:2014aa} The equilibrium partition function becomes non-analytic at a conventional phase transition as a function of the control parameter such as temperature or pressure. It turns out, that the Loschmidt amplitude can also exhibit nonanalyticities, but as a function of time rather than a control parameter. Indeed it has been shown that the rate function, which is obtained from taking the logarithm of the Loschmidt amplitude, exhibits nonanalyticities when the system is quenched across a quantum critical point whereas it remains smooth for quenches within the same dynamical phase.\cite{Heyl:2013aa, Karrasch2013, Heyl:2014aa, Andraschko2014, Kriel2014, Canovi2014, Vajna2014, Dora13, Heyl2015, Sharma2015, Zunkovic2016, Heyl2016, Goold17}. Recently, it became also possible to measure Loschmidt amplitudes in various experimental settings.~\cite{cetina_ultrafast_2016, Jurcevic16}

So far the Loschmidt amplitude has mostly been studied for one dimensional systems with discrete $\mathbb{Z}_2$ symmetries (see, however, Refs.~\onlinecite{Canovi2014, Dora13, Goold17}). In this work, we look at a three dimensional model with a continuous O(N) symmetry: the O(N) vector model. This model provides a universal description for many systems close to their critical point and is well established in the study of (non-equilibrium) quantum phase transitions.\cite{MosheJustin-Zinn2003, ChandranNanduri2013, Sciolla:2013aa, Smacchia:2015aa, ChiocchettaTavora2015, MaragaGambassi2015} For example, the equilibrium Mott-insulator to superfluid transition in the Bose-Hubbard model falls into the universality class of the O(2) model and the Heisenberg antiferromagnet can be described by an O(3) model. 

We propose the following generalization of the Loschmidt echo to systems with a continuously broken symmetry
\begin{equation}
\mathcal{L}(t) = \int\limits_{\{|\chi|=\phi_0\}} d^N\chi~|\langle \chi|\Psi(t)\rangle|^2.
\label{eq:ReturnProb}
\end{equation}
Here, $|\Psi(t)\rangle = \hat{U}(t) |\psi_0\rangle$ is the time evolved state after the quench and the integral is taken over the full set of symmetry-broken ground states $|\chi\rangle$, which can be pictured as a sphere within an $N$-dimensional space. The radius $\phi_0$ is set by the order parameter in the initial state. Below we will analyze the dynamics of the rate function associated with the Loschmidt echo 
\begin{equation}
\mathcal{R}(t) = -\frac{1}{L^d N} \log \mathcal{L}(t),
\label{eq:Rate}
\end{equation}
which shows nonanalytic behavior for quantum quenches from the dynamically ordered to the disordered phase. 

\section{The O(N) model far from equilibrium}
\label{sec:ONmodel}

The quantum O(N) model consists of $N$ real scalar fields $\hat{\Phi}_a$, $a= 1, \dots, N$ and conjugate momenta $\hat{\Pi}_a$ in $d$ spatial dimensions. The corresponding Hamiltonian is 
\begin{equation}
\hat{H} = \int_x \left[\frac{1}{2}\hat{\Pi}_a^2 + \frac{1}{2} (\nabla \hat{\Phi}_a)^2 + \frac{r_0}{2} \hat{\Phi}_a^2 + \frac{\lambda}{4!N} (\hat{\Phi}_a\hat{\Phi}_a)^2\right],
\label{eq:H}
\end{equation}
where $r_0$ is the square of the bare mass and $\lambda$ is the interaction strength. The fields obey the canonical commutation relation $[\hat{\Phi}_a(x), \hat{\Pi}_b(x')] = i \delta_{ab} \delta(x-x')$. We assume, that repeated indices are summed over. 

In the following, we consider the limit of infinitely many scalar fields, $N\rightarrow \infty$. In that limit, the interaction of strength $\lambda$ solely renormalizes the bare mass $r_0$ as follows
\begin{equation}
r = r_0 + \frac{\lambda}{6N} \langle \hat{\Phi}_a^2\rangle.
\label{eq:meff}
\end{equation}
The large-$N$ approximation relies on the factorization of the expectation value $\langle \frac{\hat{\Phi}_a\hat{\Phi}_a}{N} \hat{\Phi}_b\rangle = \langle \frac{\hat{\Phi}_a\hat{\Phi}_a}{N}\rangle\langle \hat{\Phi}_b\rangle + \mathcal{O}(1/N)$ to leading order in $1/N$.\cite{MosheJustin-Zinn2003} Therefore, there are no interactions between excitations and the model possesses an infinite number of conserved quantities and is non-ergodic.\cite{ChandranNanduri2013} As a consequence it does not thermalize. Only next-to-leading order terms introduce scattering between quasi-particle excitations and may ultimately enable thermalization.\cite{Aarts2002, Berges2002a, Weidinger17} In the present work we are not interested in the late-time thermalization physics, but rather in the transient prethermal regime after the quench; accordingly a leading order analysis is sufficient. 

In equilibrium, the O(N) model hosts two different phases: a disordered phase with finite effective mass $r>0$ and an ordered phase, in which the system spontaneously breaks the continuous O(N) symmetry by developing a finite order parameter $\langle\Phi\rangle \neq 0$. In the ordered phase the mass gap vanishes $r=0$. The equilibrium critical point is given by $r_\text{c}^\mathrm{eq} = - \frac{\lambda}{12} \int_p \frac{1}{|p|}$, which is finite for $d>1$. In $d>2$ the ordered phase extends to finite temperatures. In the rest of the paper we will focus on three spatial dimensions, $d=3$.

\begin{figure}
\includegraphics[width=.42\textwidth]{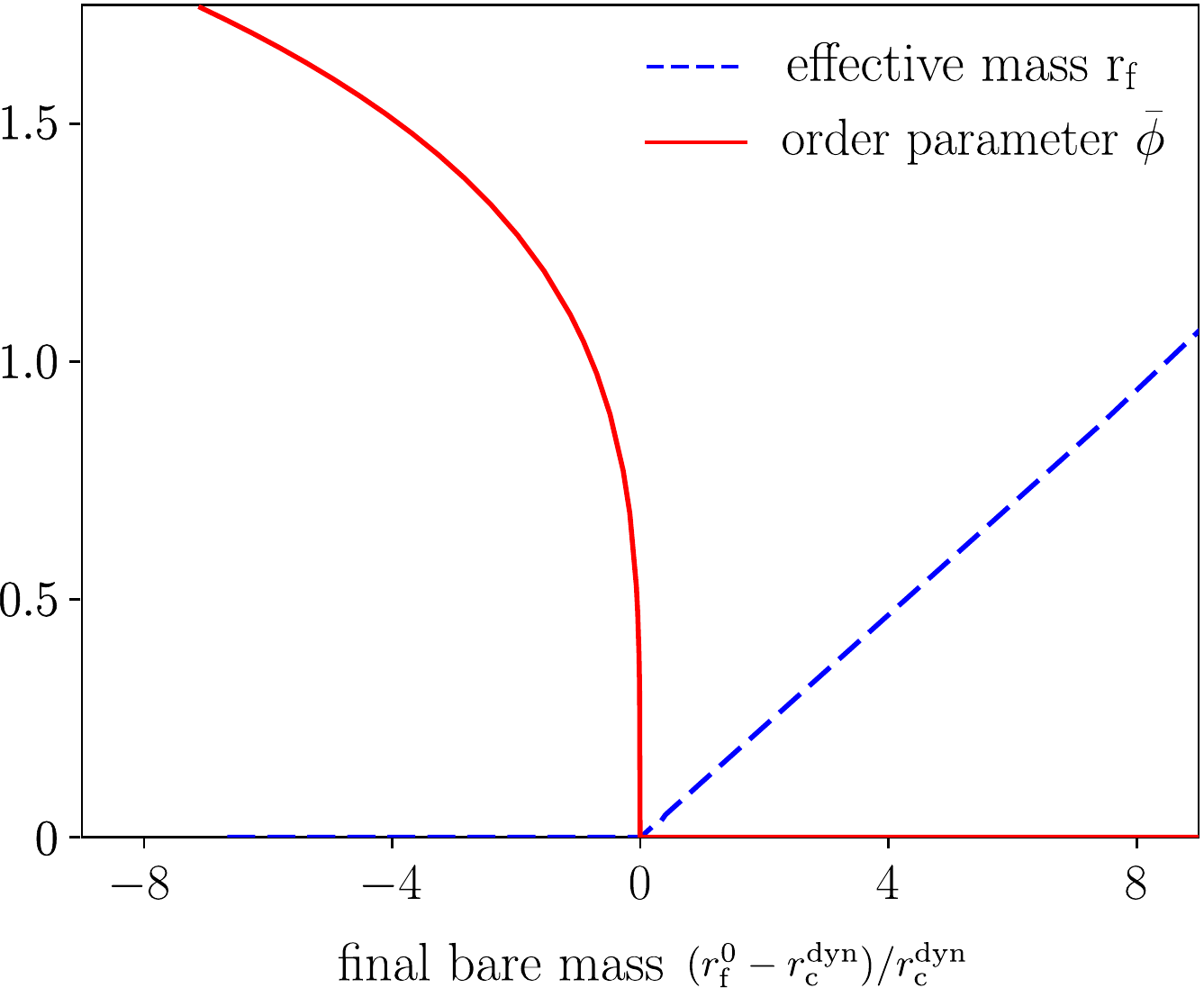}
\caption{\textbf{Dynamical phase diagram of the O(N) model in three spatial dimensions.} The system is prepared in the equilibrium symmetry broken phase at zero temperature. For quenches to a point inside of the dynamically symmetry-broken phase, $r_{\mathrm{f}}^0 < r_\mathrm{c}^\text{dyn}$, the order parameter relaxes to a finite value $\bar\phi$ (red line) and the effective mass $r_\mathrm{f}$ remains zero, indicating the presence of gapless excitations in the steady state. For quenches into the symmetric phase, $r_{\mathrm{f}}^0 > r_\mathrm{c}^\text{dyn}$, the long-time average of the order parameter is zero $\bar\phi=0$ and the effective mass $r_\mathrm{f}$ becomes finite (blue dashed line). Close to the critical point $r_\mathrm{c}^\text{dyn}$ the long-time average $\bar\phi$ vanishes as $(r_{\mathrm{c}}^\text{dyn}-r_{\mathrm{f}}^0)^{1/4}$ and the effective mass as $r_\mathrm{f} \sim (r_{\mathrm{f}}^0-r_{\mathrm{c}}^\text{dyn})$.\cite{Sciolla:2013aa}}
\label{fig:phases}
\end{figure}

Let us assume in the following, that the system has been prepared in the symmetry-broken ground state $|\Psi_0\rangle$ at $r_\mathrm{i}^{0}$, with the order parameter $\langle \hat{\Phi}_a \rangle = \delta_{1,a} \phi_0$ pointing along the $a=1$ direction. The value of $\phi_0$ is given by
\begin{equation}
(\phi_0)^2 = -\frac{6 r_\mathrm{i}^{0}}{\lambda} - (N-1) \langle\hat{\Phi}_2\hat{\Phi}_2\rangle,
\label{eq:phi0}
\end{equation}
which follows directly from the initial mass being zero. Here, we also used, that there is a remaining $O(N-1)$-symmetry for the $a\geq 2$ components. We then suddenly change the mass to the final value $r_0 = r_\mathrm{f}^{0}$ and let the system evolve in time. If the final value $r_\mathrm{f}^{0}$ is smaller than the dynamical critical value $r_{\mathrm{c}}^\mathrm{dyn}$, the system reaches an ordered steady state characterized by $r_\text{f} =0$ and $\bar{\phi} = \lim_{T\rightarrow\infty} \frac{1}{T}\int_0^\infty dt \phi(t) > 0$.\cite{Sciolla:2013aa} On the other hand, if $r_\mathrm{f}^{0} > r_{\mathrm{c}}^\mathrm{dyn}$ the order is melted. Therefore, the effective mass $r_\mathrm{f} > 0$ and the order parameter $\bar{\phi} = 0$, as illustrated in the dynamical phase diagram for $d=3$ in Fig.~\ref{fig:phases}.

To obtain  the equations of motion at $N\rightarrow \infty$, we treat the $a=1$ component of the field as a classical variable, $\hat{\Phi}_1(t) \rightarrow \phi(t)\in \mathbb{R}$, and expand the $a\geq2$ components into creation and annihilation operators that diagonalize the initial Hamiltonian\cite{Smacchia:2015aa}
\begin{equation}
\hat{\Phi}_{a\geq 2}(p, t) = f_p(t) \hat{b}^{(a)}_{p} + f_p^\ast(t) \hat{b}^{(a)\dagger}_{-p},
\label{eq:fielddecomp} 
\end{equation}
where $\hat{\Phi}_a(x, t) = V^{-1/2} \sum_p \hat{\Phi}_a(p, t) e^{ipx}$.
Note, that due to the $O(N-1)$ symmetry of the remaining $a\geq 2$ components, the time dependence is identical for all of them and hence the mode functions $f_p(t)$ in Eq.~\eqref{eq:fielddecomp} do not carry a field component index.

Using the Heisenberg equations of motions, we obtain
\begin{align}
&\ddot{f}_p(t) + [p^2 + r(t)] f_p(t) = 0 \notag\\
&\ddot{\phi}(t) + r(t) \phi(t) = 0,
\label{eq:EOM}
\end{align}
with the time-dependent effective mass
\begin{equation}
r(t) = r_{\mathrm{f}}^0 + \frac{\lambda}{6N}\left(\phi^2(t) + (N-1)\int_p |f_p(t)|^2\right).
\label{eq:rt}
\end{equation}
It is important to notice, that $\phi(t)\sim \sqrt{N}$. Therefore, both terms in the parenthesis in Eq.~\eqref{eq:rt} scale linearly with $N$ and contribute to the effective mass.

The initial conditions of Eq.'s~\eqref{eq:EOM} are $f_p(0) = 1/\sqrt{2|p|}$, $\dot{f}_p(0) = -i\sqrt{|p|/2}$, which follow from requiring that $\hat{b}_p$, $\hat{b}^\dagger_p$ diagonalize the initial Hamiltonian and $r(t=0) = 0$. Furthermore we have $\phi(0) = \phi_0$ and $\dot{\phi}(0) = 0$, with $\phi_0$ given by Eq.~\eqref{eq:phi0}. To regularize the infrared divergence of $f_p(0)$, we introduce an  cut-off $p_0 = 2\pi/L$, with $L$ being the linear extension of the system. This amounts to placing the field theory in a finite box with volume $L^d$. Eventual UV divergencies are regularized with a finite cut-off $\Lambda$ in momentum space.

\section{Results}
\subsection{Time evolved state}
\label{sec:state}
In order to calculate the return probability to the groundstate manifold, we need to know the time evolved state $|\Psi(t)\rangle = \hat{U}(t)|\Psi_0\rangle$. In the $N\rightarrow\infty$ limit the state $|\Psi(t)\rangle$ factorizes in the field components due to the effectively quadratic Hamiltonian at leading order.\cite{Smacchia:2015aa} In the $a\geq2$ components there is a squeezed state $|\psi_\mathrm{sq}(t)\rangle$ and in the "classical" $a=1$ component a coherent state $|\phi(t)\rangle$ ,
\begin{align}
&|\Psi(t)\rangle = |\phi(t)\rangle \otimes |\psi_{\mathrm{sq}}(t)\rangle \notag\\
&|\phi(t)\rangle = e^{-\frac{1}{2} \gamma^2 \phi^2(t)}e^{ \gamma  \phi(t) \hat{b}_{p_0}^{(1)\dagger}} |0\rangle \notag\\
&|\psi_{\mathrm{sq}}(t)\rangle = \prod_{\substack{p>0 \\ a\geq 2}} \frac{1}{\sqrt{|\alpha_p(t)|}} \exp\left\{\frac{\beta_p^\ast(t)}{2\alpha_p^\ast(t)} (\hat{b}_{p}^{(a)\dagger})^2\right\}|0\rangle,
\label{eq:state}
\end{align}
where $\alpha_p(t) = f_p(t) \sqrt{\frac{|p|}{2}} + i \frac{\dot{f}_p(t)}{\sqrt{2|p|}}$, $\beta_p(t) = f_p(t) \sqrt{\frac{|p|}{2}} - i \frac{\dot{f}_p(t)}{\sqrt{2|p|}}$ and $\gamma = L^{\frac{d-1}{2}} \left(\frac{\sqrt{d}\pi}{2}\right)^\frac{1}{2}$. The coherent state contribution gives rise to a finite order parameter $\langle \Psi(t)|\hat{\Phi}_1|\Psi(t)\rangle = \phi(t)$.

\subsection{Return probability to the groundstate manifold}
\label{sec:returnprob}
An arbitrary state in the groundstate manifold of~\eqref{eq:H} in the symmetry-broken phase can be written as 
\begin{equation}
|\chi\rangle = e^{-\frac{1}{2} \gamma^2 \chi^2}e^{ \gamma \chi^T \hat{b}_{p_0}^{\dagger}}|0\rangle,
\label{eq:initstate}
\end{equation}
where $\chi = (\chi_1, \dots, \chi_N)$, $|\chi|^2 = \phi_0^2$ and $\hat{b}_p = (\hat{b}^{(1)}_p, \dots, \hat{b}^{(N)}_p)$. The expectation of the field-operator in this state is given by $\langle \chi|\hat{\Phi}_a|\chi\rangle = \chi_a$.   
The overlap $\langle\chi|\Psi(t)\rangle$ factorizes into a product over the field components. For $a=1$ we get a scalar product of two coherent states and for $a\geq2$ we have scalar products of a coherent and a squeezed state, which we calculate by expanding the exponentials. For the return probability to a specific initial state, we obtain
\begin{align}
|\langle &\chi|\Psi(t)\rangle|^2 =  \exp\Biggl\{-L^d N \int_p \log |\alpha_p(t)| \notag\\
&-L^{d-1} \frac{\sqrt{d}\pi}{2} [\phi^2(t) + \phi^2(0) - 2 \chi_1 \phi(t) + \sum\limits_{a\geq 2} \chi_a^2]\Biggr\}.
\label{eq:overlap1}
\end{align}
In deriving this formula we also made use of the fact, that for large systems, $L\gg 1$, i.e., small $p_0=2\pi/L$, the ratio $\beta_{p_0}(t)/\alpha_{p_0}(t)$ approaches $1$.  
\begin{figure}
	\includegraphics[width=.44\textwidth]{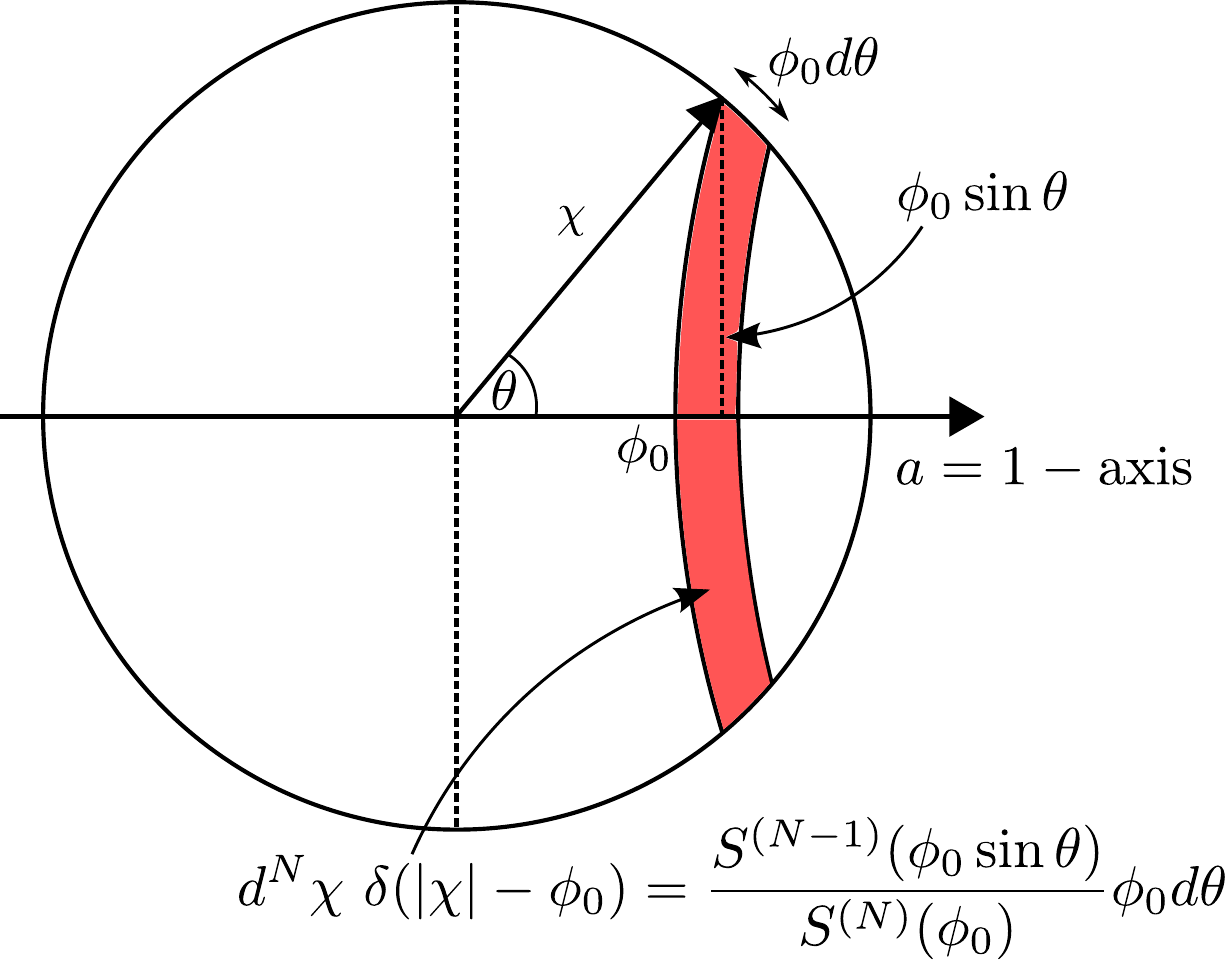}
	\caption{\textbf{Definition of the angle} $\bf\theta$ \textbf{and surface of element of the N-dimensional sphere.} The ground state manifold of the $O(N)$ model can be pictured as a sphere with radius $\phi_0$ in a $N$-dimensional space. The return probability $\mathcal{L}(t)$ to the ground state manifold is obtained from the integral of the overlap $|\langle \chi|\Psi(t)\rangle|^2$ over this sphere. Defining $\theta$ as the angle between the vector $\chi$ and the initial order parameter $(\phi_0, 0, \dots, 0)$ and making use of the rotational symmetry around the $(a=1)$-axis, one can write the integration element $d^N \chi~\delta(|\chi|-\phi_0)$ as the product of the arc length $\phi_0 d\theta$ and the surface area of the sphere in $N-1$ dimensions $S^{(N-1)}(\phi_0 \sin\theta)$ generated by rotating $\chi$ around the $a=1$ - axis with $\theta$ fixed. To obtain a probability measure, we finally divide the integration element by the total available surface area $S^{N}(\phi_0)$.}
	\label{fig:sphere}
\end{figure}
\begin{figure*}
	\includegraphics[width=.98\textwidth]{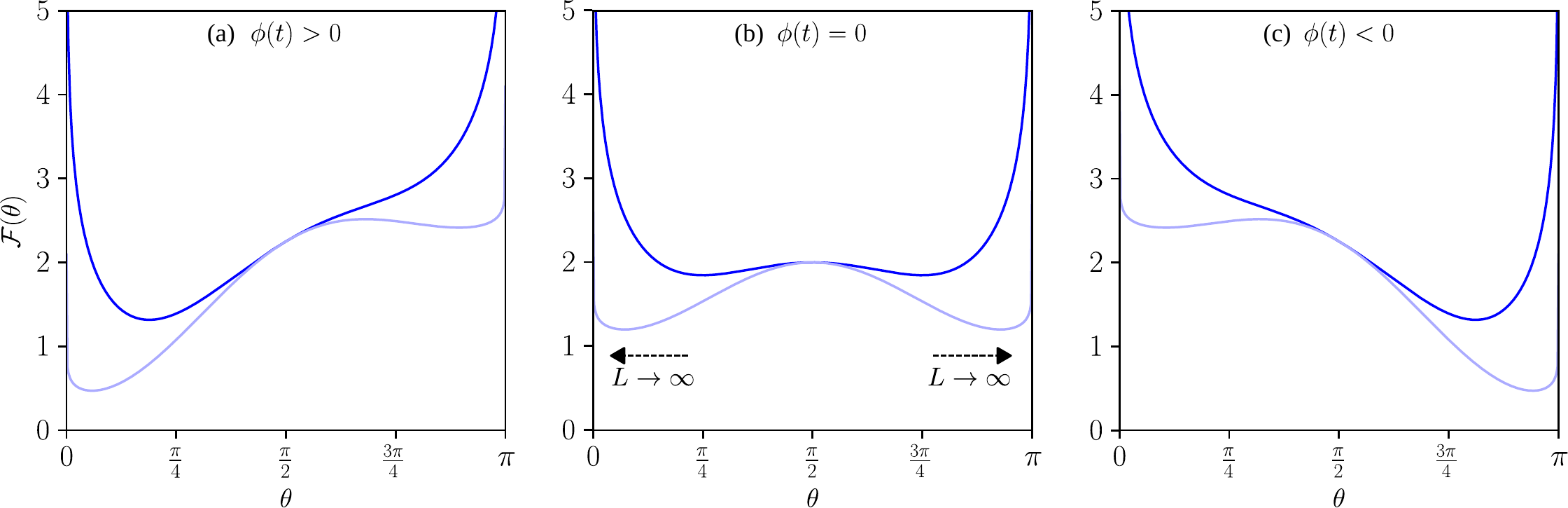}
	\caption{\textbf{Order parameter landscape for the angle $\theta$:} The return probability, \eq{eq:ThetaPartition}, can be interpreted as a classical partition function for the variable $\theta\in[0, \pi]$ moving in an effective free energy landscape $\mathcal{F}(\theta, \phi)$. The landscape has the form of a double well potential, where the order parameter $\phi(t)$ is acting as an external field, shifting the two wells against each other. For \fc{a} $\phi(t)>0$ the left minimum is energetically more favorable, while for \fc{c} $\phi(t)<0$ the situation is reversed. As the system size $L$ is increased, the left (right) potential well is shifted toward $\theta=0$ ($\theta=\pi$), which correspond to the states having an order parameter parallel  (anti-parallel) to the initial state. \fc{b} When $\phi(t)$ changes sign the most relevant value of $\theta$ jumps from one well to the other, which gives rise to the kinks in the Loschmidt rate function $\mathcal{R}(t)$.}
	\label{fig:ThetaEnergy}
\end{figure*}

The overlap $|\langle \chi|\Psi(t)\rangle|^2$ is rotational invariant around the $a=1$ axis. Hence, we use spherical coordinates (see \fig{fig:sphere}) to calculate the integral over the groundstate manifold as required in Eq.~\eqref{eq:ReturnProb}. Defining $\theta \in [0, \pi]$ as the angle between the vector $\chi$ and the $a=1$ axis, i.e., $\cos \theta = \chi_1/\phi_0$, we can write 
\begin{align}
|\langle \chi&|\Psi(t)\rangle|^2 = \mathcal{N}_{\mathrm{sq}}(t)\exp\Biggl\{-L^{d-1} \frac{\sqrt{d}\pi}{2} \phi^2(0) \notag\\
&\times\left[1+\left(\frac{\phi(t)}{\phi(0)}\right)^2 + \sin^2 \theta - 2 \left(\frac{\phi(t)}{\phi(0)}\right) \cos \theta \right]\Biggr\}.
\end{align}
Here, we introduced the abbreviation $\mathcal{N}_\mathrm{sq}(t) = \exp[-L^d N \int_p \log |\alpha_p(t)| ]$.
The integration element can be written as $d^N \chi~\delta(|\chi| - \phi_0)= S^{N-1}(\phi_0 \sin \theta) \phi_0 d \theta / S^{N}(\phi_0)$, where $S^{n}(r)=2\pi^{\frac{n}{2}}\Gamma(n/2)^{-1} r^{n-1}$ is the surface of the n-sphere (see \fig{fig:sphere} for a graphical interpretation). Exponentiating the $\sin \theta$ - term, we obtain the return probability to the ground state manifold
\begin{equation}
\mathcal{L}(t) = A \, \mathcal{N}_{\mathrm{sq}}(t)\int\limits_0^\pi d\theta e^{-L^{d-1}N \mathcal{F}(\theta, \phi(t))},
\label{eq:ThetaPartition}
\end{equation}
with 
\begin{align}
\mathcal{F}(\theta, \phi) =  \sqrt{\frac{\pi}{2}} \frac{\phi_0^2}{N} &\left[1+\left(\frac{\phi}{\phi_0}\right)^2 - 2\left(\frac{\phi}{\phi_0}\right) \cos \theta + \sin^2\theta \right] \notag\\
&- \frac{N-2}{N} L^{-d+1} \log \sin \theta
\label{eq:ThetaEnergy}
\end{align}
and a constant $A={\pi}^{-1/2} \Gamma(\frac{N}{2})/\Gamma(\frac{N-1}{2})$. We will refer to $\mathcal{L}(t)$ also as Loschmidt Echo. Eq.~\eqref{eq:ThetaPartition} can be interpreted as a classical partition function of the angular variable $\theta$ moving in an order parameter landscape $\mathcal{F}(\theta, \phi(t))$, with $L^{d-1}N$ playing the role of inverse temperature. The energy landscape, Eq.~\eqref{eq:ThetaEnergy}, has the shape of a double well potential, where the order parameter $\phi$ is acting as an external field tilting the two wells against each other, see Fig.~\ref{fig:ThetaEnergy}. The two wells are energetically equivalent, when the external field vanishes ($\phi(t)=0$).The larger $L$, the more the two wells move outwards to $0$ and $\pi$ . Nevertheless, the $\log\sin\theta$ term is important, because it is responsible for creating the double well landscape. 

In the thermodynamic limit $L\gg1$, we can evaluate the integral in Eq.~\eqref{eq:ThetaPartition} using a saddle point approximation. Taking this limit corresponds to very low temperatures in the classical partition function and the variable $\theta$ will pick the minimum energy well
\begin{equation}
\mathcal{L}(t)\underset{L\gg 1}{\simeq} \mathcal{N}_\mathrm{sq}(t) \exp\left[-L^{d-1} N \min_{\theta\in [0, \pi]} \mathcal{F}(\theta, \phi(t))\right].
\label{eq:returnProbResult}
\end{equation} 
For $L \to \infty$, the last term in Eq.~\eqref{eq:ThetaEnergy} vanishes and the minimum is at $\theta_\mathrm{min} = 0$ ($\theta_\mathrm{min} = \pi$) for $\phi(t)>0$ ($\phi(t)<0$), meaning that $\chi$ is parallel (antiparallel) to the order parameter of the initial state. Therefore, only two states from the continuous ground state manifold contribute significantly to the Loschmidt echo: $\mathcal{L}(t) \sim (|\langle +\phi_0|\Psi(t)\rangle|^2+|\langle -\phi_0|\Psi(t)\rangle|^2)$. This can be interpreted as follows: the order-parameter oscillates only along a fixed axis due to the symmetry of the Hamiltonian and cannot explore the whole ground state manifold.

Our result for the coherent state contribution to the Loschmidt rate function scales subextensively with system size as $\sim L^{d-1}$, see the prefactor of $\mathcal{F}(\theta, \phi(t))$ in \eq{eq:returnProbResult}. This is a consequence of the infrared divergence of the initial mode function $f_p(0)$ due to the spontaneously broken symmetry, which leads to the scaling of $\gamma \sim L^{(d-1)/2}$ in the coherent state, \eq{eq:state}. From that, the wavefunction overlap $\langle \chi|\Psi(t)\rangle$ of the time evolved state and an arbitrary state in the ground state manifold contains terms, that scale subextensively $\sim L^{d-1}$. We emphasize that the subextensive scaling shows up only in the wave function overlap but not  in  expectation values of observables. Examples include the order parameter and the work performed in a quench. The latter shows a normal extensive scaling $\sim L^d$ with system size. The average work $\langle \hat{H}_f\rangle$ is given by the expectation value of the post-quench Hamiltonian in the initial state, $\langle \hat{H}_f\rangle = L^d \frac{r_\mathrm{f}^{0}-r_\mathrm{i}^{0}}{2} (N\int_p\frac{1}{2|p|} + \phi_0^2)$. All higher cumulants of the work distribution function vanish in our leading order approximation. Generally, the logarithm of the Loschmidt amplitude acts as the generating function for cumulants of the work-distribution.\cite{Talkner07, Silva08, Campisi11} We also find in our model that to leading order in $N$, the Loschmidt echo reproduces exactly the cumulants of the work.

\begin{figure*}
	\includegraphics[width=.98\textwidth]{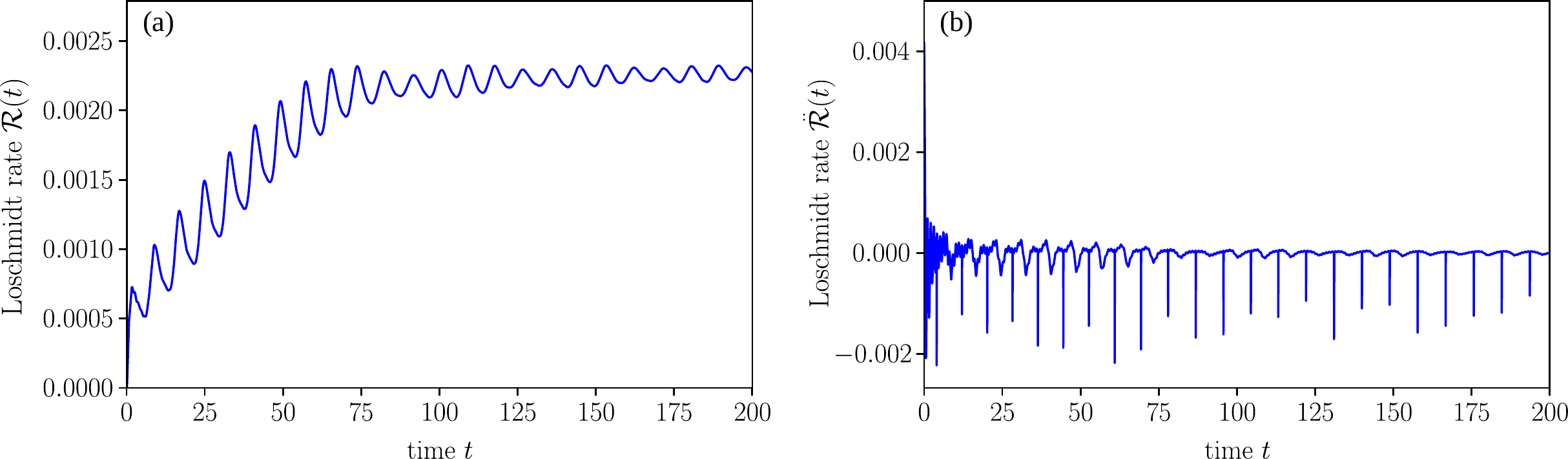}
	\caption{\textbf{Loschmidt rate function.} The Loschmidt rate function $\mathcal{R}(t)$ of the return probability to the groundstate manifold contains a squeezed state contribution $\mathcal{R}_\mathrm{sq}(t)$, which scales extensively with system size, and a coherent state contribution $\mathcal{R}_\mathrm{coh}(t)$ that scales subextensively. Whereas $\mathcal{R}_\mathrm{sq}(t)$ is a smooth function of time, $\mathcal{R}_\mathrm{coh}(t)$ shows kinks, when the system is quenched across the dynamical quantum phase transition. Due to the subextensive scaling of the coherent state contribution $\mathcal{R}_\mathrm{coh}$, the non-analytic behavior is not visible in the full rate-function $\mathcal{R}(t)$ of the return probability, (a). Nevertheless the non-analytic behavior is clearly observable in the second derivative $\ddot{\mathcal{R}}(t)$ as $\delta$-peaks, since the squeezed state contribution relaxes on a much shorter time scale than the one of the coherent state, (b). The system parameters are $L=2.5\times 10^4$, $\lambda = 1.0$, $r_\mathrm{i}^{0}=-1.0$ and $r_\mathrm{f}^{0}=0.0$.}
	\label{fig:lorate}
\end{figure*}
\subsection{Rate function}
\label{sec:ratefunction}
Calculating the rate function $\mathcal{R}(t) = -L^{-d}N^{-1}\log\mathcal{L}(t)$ from Eq.\eqref{eq:returnProbResult}, we find that
\begin{align}
&\mathcal{R}(t) = \mathcal{R}_\mathrm{sq}(t) + \frac{1}{L} \mathcal{R}_\mathrm{coh}(t)\notag\\
&\mathcal{R}_\mathrm{sq}(t) = \int_p \log|\alpha_p(t)|\notag\\
&\mathcal{R}_\mathrm{coh}(t) = \sqrt{\frac{\pi}{2}} N^{-1} \phi_0^2 \left[1+\left(\frac{\phi(t)}{\phi_0}\right)^2 - 2\left|\frac{\phi(t)}{\phi_0}\right|\right].
\label{eq:RateResult}
\end{align}
The contribution from the squeezed state $\mathcal{R}_\mathrm{sq}$ is obtained from $\mathcal{N}_\mathrm{sq}$, and the coherent state contribution is obtained by explicitly calculating the minimum in Eq.~\eqref{eq:returnProbResult}. The rate function $\mathcal{R}_\mathrm{sq}$ is a smooth function of time, since $|\alpha_p(t)|$ is smooth and bounded from below by 1. $\mathcal{R}_\mathrm{coh}$ on the other hand exhibits kinks at zero crossings of $\phi(t)$ due to the absolute value in the last term of Eq.~\eqref{eq:RateResult}. As discussed above, the coherent state contribution is suppressed by a factor of $L^{-1}$. 
However, the squeezed-state part of the rate function $\mathcal{R}_\mathrm{sq}(t)$ relaxes to a constant value on a much shorter time-scale than the order parameter $\phi(t)$, because of an integral over momenta. Therefore, the non-analyticities in $\mathcal{R}_\mathrm{coh}(t)$ can be identified for instance in the second derivative $\ddot{\mathcal{R}}(t)$ of the rate function. For the squeezed state, $\ddot{\mathcal{R}}_\mathrm{sq}(t) \approx 0$, whereas the coherent state retains prominent $\delta$-peaks $\ddot{\mathcal{R}}_\mathrm{coh}(t) \sim \sum_{T_\mathrm{kink}} \delta(t - T_\mathrm{kink})$, as illustrated in Fig.~\ref{fig:lorate}.

\begin{figure}
\includegraphics[width=.44\textwidth]{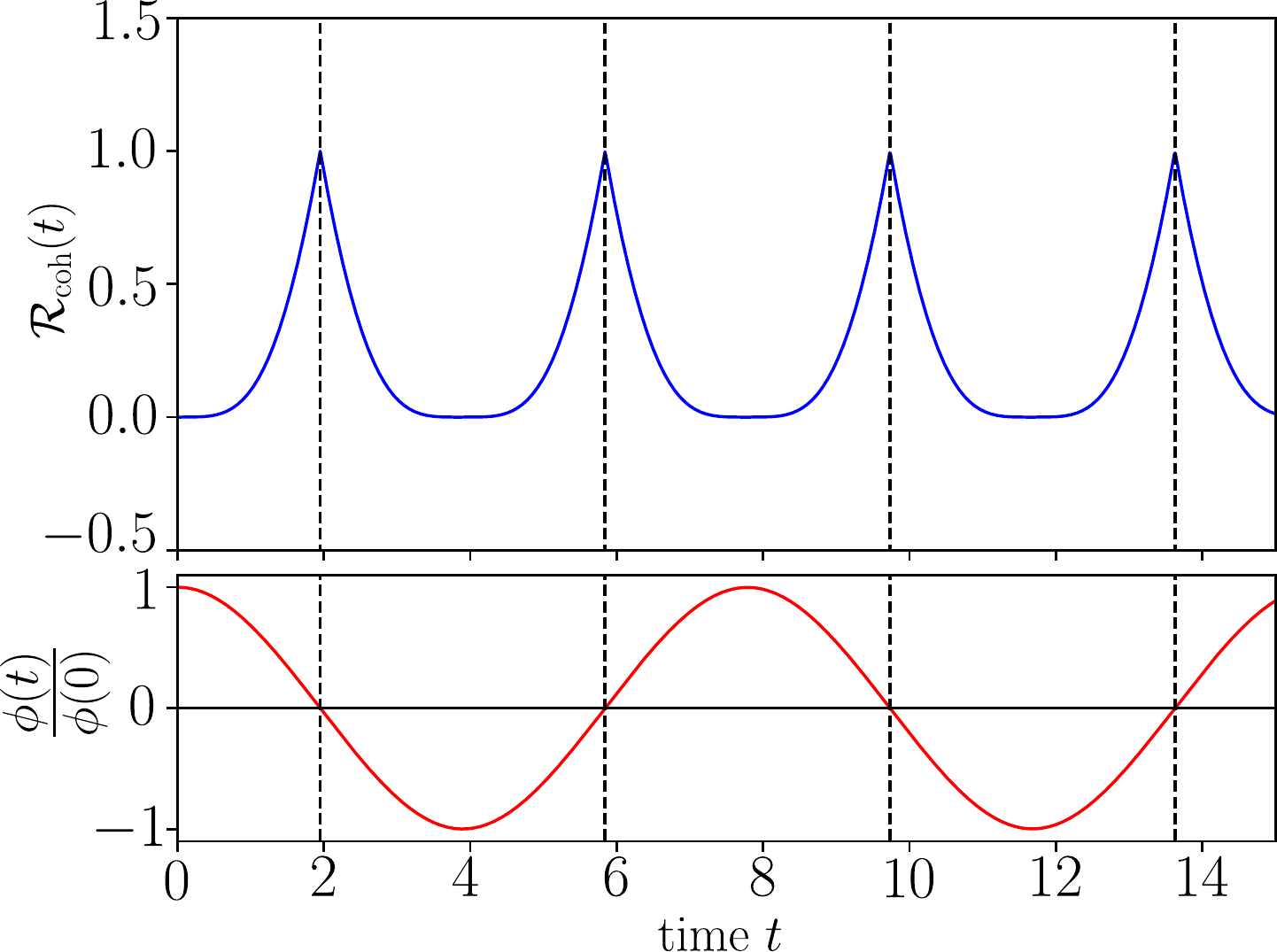}
\caption{\textbf{Coherent state contribution to the rate function and order parameter dynamics.} The coherent state contribution $\mathcal{R}_\mathrm{coh}$ to the rate function exhibits kinks at the zero-crossings of the order parameter $\phi(t)$. The kinks appear periodically and the time between them $\Delta T_\mathrm{kink}$ is determined by final effective mass $r_\mathrm{f}$: $\Delta T_\mathrm{kink} \sim (r_\mathrm{f}^{0} - r_{\mathrm{c}}^\mathrm{dyn})^{-\frac{1}{2}}$. The data is evaluated for the same parameters as in \fig{fig:lorate} }
\label{fig:coherent_rate}
\end{figure}

The coherent state contribution to the Loschmidt rate function $\mathcal{R}_\mathrm{coh}(t)$ exhibits kinks at the zero crossings of the order parameter, $\phi(T_\mathrm{kink})=0$, see Fig.~\ref{fig:coherent_rate}. From the numerical solution of the equations of motion~\eqref{eq:EOM} we also find that the order parameter relaxes to a non-zero value for quenches inside the dynamical symmetry-broken phase ($r_\mathrm{f}^{0} < r_{\mathrm{c}}^\mathrm{dyn}$). In this case there are no zero crossings of $\phi(t)$ and hence we do not find any non-analyticities in $\mathcal{R}_\mathrm{coh}$. By contrast, for quenches to the symmetric phase ($r_\mathrm{f}^{0} > r_{\mathrm{c}}^\mathrm{dyn}$), the order parameter oscillates around zero and approaches $\bar\phi=0$ and $\mathcal{R}_\mathrm{coh}$ exhibits kinks. As a consequence, there is an intimate relation between the dynamical phase transition of the order parameter and the kinks in the Loschmidt rate function of the return probability to the groundstate manifold.

Following a quench to the symmetric phase, the effective mass $r(t)$, Eq.~\eqref{eq:rt}, attains a finite average value $r_\mathrm{f}$, which feeds back into the equations of motion, Eq.~\eqref{eq:EOM}, as frequency squared of $\phi(t)$. Accordingly, the kinks in $\mathcal{R}_\mathrm{coh}$ appear at equidistantly spaced times $T_\mathrm{kink}$ and the time $\Delta T_\mathrm{kink}$ between two kinks is uniquely determined by $r_\mathrm{f}$. The effective mass after a quench to the symmetric phase scales linearly with the distance of the final bare mass $r_\mathrm{f}^{0}$ from the dynamical critical point $r_{\mathrm{c}}^\mathrm{dyn}$, $r_f \sim r_\mathrm{f}^{0} - r_{\mathrm{c}}^\mathrm{dyn}$, as depicted in Fig.~\ref{fig:phases}. We therefore find
\begin{equation}
\Delta T_\mathrm{kink} = \frac{\pi}{\sqrt{r_\mathrm{f}}} \sim (r_\mathrm{f}^{0} - r_{\mathrm{c}}^\mathrm{dyn})^{-\frac{1}{2}}.
\label{eq:DeltaT}
\end{equation}
Therefore, the time between the kinks diverges with the same critical exponent upon approaching the dynamical critical point as the correlation length in equilibrium, which is a manifestation of the O(N) model being a relativistic field theory in which time and space scale in the same way.

\section{Conclusion and Outlook \label{sec:outlook}}

We have studied the rate function of the return probability to the ground state manifold in the O(N) model following a quantum quench from a symmetry breaking initial state to the symmetric phase. The rate function exhibits kinks, which are located at the zero crossings of the order parameter $\phi(t)$ and are equally spaced with a period $\Delta T_\mathrm{kink}$ determined by the final effective mass. In our model, the non-analytic contribution to the return probability scales subextensively with system size. Such a subextensive contribution can also appear in equilibrium whenever a system undergoes a surface or impurity phase transition.

For quenches from the symmetric to the symmetry-broken phase kinks are absent, since the closing of the gap leads to a divergent time scale between kinks. Also, due to the absence of explicit symmetry-breaking terms in the Hamiltonian, no finite order parameter can be ever generated. 

Our results for the non-equilibrium dynamics are obtained fully analytically to leading order in the number of components $N$ of the field theory. We point out that the saddlepoint approximation, which we employ in the calculation of the return probability, only relies on the thermodynamic limit $L\to \infty$ and not on $N$ being large. Furthermore the presence of kinks in the rate function $\mathcal{R}(t)$ hinges on the presence of the coherent state, i.e., a finite order parameter $\phi(t)$. Next-to-leading order corrections would modify the time evolution of the order parameter and the quantum fluctuations in the time evolved state, but would not destroy the symmetry-broken phase, i.e. the coherent contribution to the time evolved state. Therefore, we argue, that our results remain valid beyond a leading order approximation in $1/N$. Moreover, due to the universality of the O(N) model, we expect our results to be generic for dynamical critical points in models with continuous symmetries. In particular, the return probability should be  dominated by the states parallel and anti-parallel to the initial state, leading to non-analytic behavior of the rate function for quenches from the symmetry-broken to the symmetric phase. Moreover, the zero crossings of the order parameter should determine the times at which nonanalyticities appear in the Loschmidt echo. It would be intriguing to explore these findings in other models with continuous symmetry breaking.

\begin{acknowledgments}
\textit{\textbf{Acknowledgments.---}}We acknowledge support from the Deutsche Forschungsgemeinschaft via the Gottfried Wilhelm Leibniz Prize program (MH), the Technical University of Munich - Institute for Advanced Study, funded by the German Excellence Initiative and the European Union FP7 under grant agreement 291763 (SW, MK), and from the DFG grant No. KN 1254/1-1 (SW, MK).
\end{acknowledgments}

\bibliography{Loschmidt_arxiv_v11.bib}

\end{document}